# Dark Matter Interpretation of Neutron Multiplicity Anomalies

Thomas Ward[a,b] and Wladyslaw H. Trzaska[c,d]

[a] *US Department of Energy, Office of Nuclear Energy, 1000 Independence Ave., Washington D.C., USA*

[b]*Techsource Incorporated, 1415 Central Ave., Suite 210, Los Alamos, New Mexico, USA*

[c]*Department of Physics, University of Jyväskylä, P.O. Box 35, FI-40014 University of Jyväskylä, Finland*

[d]*Helsinki Institute of Physics (HIP), P.O. Box 64, 00014 University of Helsinki, Finland*

E-mail: tward@techsource-inc.com

Abstract. Subterrestrial neutron spectra show weak but consistent anomalies at multiplicities ~100 and above [1-3]. The data of the available measurements are of low statistical significance [4] but indicate an excess of events not correlated with the muon flux. The origin of the anomalies remains ambiguous but could be a signature of WIMP annihilation-like interaction with a Pb target. In this paper, we outline a model consistent with this hypothesis, the extended Standard Model (SM) approach called the Radiation Gauge Model (RGM) [5]. The RGM identifies scalar neutrino-antineutrino wave function components of WIMP Dark Matter (DM) responsible for the weak interaction leading to annihilation with ordinary matter. The model assigns neutrino-nucleon(target) charged current (CC) transitions to the observed anomalies. If the existence of the anomalies is confirmed and the model interpretation is positively verified, this will be the first terrestrial indirect detection of DM.

1. **Introduction**

As it was reported at ICRC 2021 [1], TAUP 2021 [2], and VCI 2022 [3], subterrestrial neutron spectra show weak but consistent anomalies at multiplicities ~100 and above. The data of the available measurements are still of low statistical significance [4] but indicate an excess of events not correlated with the underground muon flux. The origin of the anomalies remains ambiguous but, in principle, could be a signature of a WIMP DM annihilation-like interaction with a massive Pb target. Here, we outline a model consistent with this WIMP hypothesis. We use an extended SM approach, the RGM [5] phenomenology, that identifies the scalar neutrino-antineutrino wave function component of WIMP DM responsible for the weak interaction leading to annihilation with ordinary matter. The RGM model assigns neutrino-nucleon (target) charged current (CC) transitions to the observed anomalies, i.e., $p_A + \bar{\nu}_{DM} \to n_A + l^+$ and $n_A + \nu_{DM} \to p_A + l^-$. For example, an 8 GeV WIMP particle annihilating a Pb nucleus requires 3.25 GeV excitation to fragment the Pb into neutrons and protons, which further undergo (n, xn) and (p, xn) reactions in the massive Pb target. The outgoing weak interaction leptons (e, mu, tau, and neutrinos) emit the remainder of the energy (4.75 GeV). If the existence of the anomalies is confirmed and the model interpretation is positively verified, this will be the first terrestrial indirect detection of DM.

**2.1    RGM-SM Phenomenology**.

RGM is a simple extension of the spin dependent SM, mixing the four fundamental radiation gauge boson fields, pseudoscalar (Higgs), vector (EM), axial vector (WZ) and tensor (graviton) with complimentary Yang-Mills (Y-M) fields producing four EW isospin symmetry breaking (EWSB) (2x2) mixing matrices. The four-gauge boson spin fields consisting of Weak $(J^\pi = 1^+)$, EM $(J^\pi = 1^-)$, Strong $(J^\pi = 0^-)$ and Gravitational $(J^\pi = 2^+)$ are constructed from the SM Yang-Mills-Dirac-Higgs-Yukawa Lagrangian

$$L_{SM} = L_{dirac}[\psi, A] + L_{Higgs}[\psi, A] + L_Y[\psi_L, \phi, \psi_R] + L_{YM}[A] \qquad (1)$$

and four-complementary Yang-Mills Fields, one abelian $U(1)_Y$ and three non-abelian $SU(2)_L$

$$L_{YM}[A] = -1/4 F^{A\alpha}_{\mu\nu} F^{A\mu\nu}_\alpha \qquad (2)$$

The EWSB interaction removes the degeneracy of the neutral gauge fields by off-diagonal mixing via 2x2 matrices with complementary Y-M Fields, i.e., the well-known case of the W-boson field removing the isospin degeneracy of the Z-particle with EWSB

$$M^2_{WZ} = \begin{vmatrix} W & Z^0 \\ A'_{EW} & A_{EW} \end{vmatrix}. \qquad (3)$$

The $SU(2)_L x U(1)_Y$ EWSB matrices and QCD $SU(3)_C$ Supersymmetry Breaking (SSB) [6] of the Higgs field combine to form the EWSB-SSB $SU(3)_C x SU(2)_L x U(1)_Y$ matrix,

$$M^2_{EWSB-SSB} = \begin{vmatrix} M^2_{WZ}(1^+) & M^2_{Higgs}(0^-) \\ M^2_G(2^+) & M^2_{EM}(1^-) \end{vmatrix} \otimes M^2_{Higgs}(0^+) . \qquad (4)$$

Dark Matter is identified as the massive residual Y-M tensor boson resulting from the solution of the gravity tensor matrix solution

$$M_G^2(J^\pi = 2^+) = \begin{vmatrix} G & \chi_{DM}^0 \\ g & A_{EWS} \end{vmatrix}, \tag{5}$$

with the G-field, $E_G^0 = 2m_N c^2$, final state graviton, $E_{graviton} \equiv 0.00 GeV$, and the residual Y-M massive particle or DM, $M(\chi_{DM}^0) = (7.93 \pm 0.12) GeV$. The spin-parity of particle DM is $J^\pi = 2^+$. The composition of DM with scalar pairs of leptons (54%), quarks (23%) and nucleons (23%) follows from the EWSB-SSB matrix. Scalar pairs are deeply bound within the 8 GeV DM potential, which allows only neutral and near massless neutrinos to take part in the weak interaction decay and annihilation. WIMP annihilation with baryonic matter deposits roughly half the DM particle energy (about 3-4 GeV) to the baryonic matter, annihilating it into neutrons, protons, and gamma-rays, with the remainder of energy carried by the final state leptons in the weak interaction.

**DM-Baryon Annihilation:** There are four Indirect reactions from annihilation in a Pb target. They should produce four groups of events with large neutron multiplicities. The measurements and analysis of the NMDS and NEMESIS experiments are still in progress, but they point to the possible presence of such anomalies, especially in the smoothed NMDS spectrum shown in Figure 2. The plots depict the same data analyzed by Gaussian smoothing and by summing up channels. An exponential (left curve) and a power-law background (right curve) were tried. In each case, the results are consistent with the four-peak hypothesis, albeit with low significance. The observed multiplicity is equal to the efficiency times the actual multiplicity.

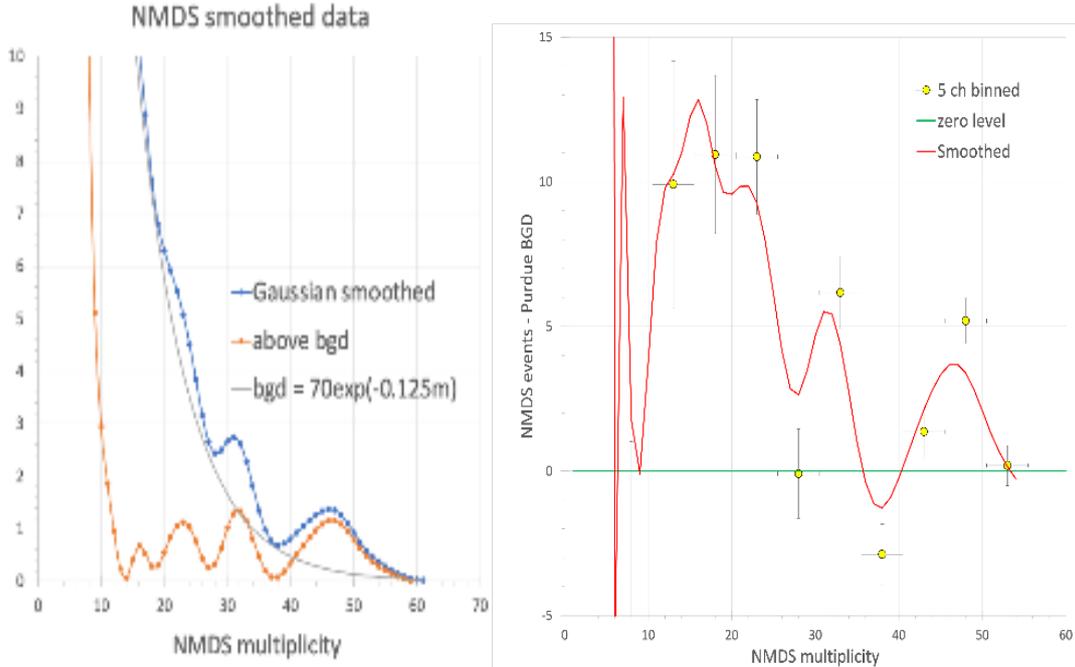

Figure 2. Smoothed (left) and binned (right) NMDS spectrum interpreted in the framework of the four anomalies explained in the text.

Indirect Inelastic Neutral Current (NC) Transitions.
(m=16.3, M=75)

$$\begin{bmatrix} \nu \\ \bar{\nu} \end{bmatrix}_{\chi^0}^{e,\mu,\tau} + \begin{bmatrix} nn \\ pp \end{bmatrix}_{Pb}^{3} \xrightarrow{Z^0}_{Z^0} \begin{bmatrix} \nu' \\ \bar{\nu}' \end{bmatrix}_{\chi^{0'}} \oplus \begin{bmatrix} n'n \\ p'p \end{bmatrix}_{Pb'}^{3}$$

Indirect Charged Current (CC) Transitions
(m=22.9, M=105)

$$\begin{bmatrix} \nu \\ \bar{\nu} \end{bmatrix}_{\chi^0}^{e,\mu,\tau} + \begin{bmatrix} nn \\ pp \end{bmatrix}_{Pb}^{3} \xrightarrow{Z^0}_{W^+} \begin{bmatrix} \nu \\ l^+ \end{bmatrix}_{\chi^0} \oplus \begin{bmatrix} nn \\ np \end{bmatrix}_{Pb}^{3}$$

(m=32.5, M=149)

$$\begin{bmatrix} \nu \\ \bar{\nu} \end{bmatrix}_{\chi^0}^{e,\mu,\tau} + \begin{bmatrix} nn \\ pp \end{bmatrix}_{Pb}^{3} \xrightarrow{W^-}_{Z^0} \begin{bmatrix} l^- \\ \bar{\nu} \end{bmatrix}_{\chi^0} \oplus \begin{bmatrix} pn \\ pp \end{bmatrix}_{Pb}^{3}$$

(m=46.9, M=215)

$$\begin{bmatrix} \nu \\ \bar{\nu} \end{bmatrix}_{\chi^0}^{e,\mu,\tau} + \begin{bmatrix} nn \\ pp \end{bmatrix}_{Pb}^{3} \xrightarrow{W^-}_{W^+} \begin{bmatrix} l^- \\ l^+ \end{bmatrix}_{\chi^0} \oplus \begin{bmatrix} p'n \\ n'p \end{bmatrix}_{Pb'}^{3}$$

The RGM spin-dependent (SD) cross-section projections are shown in Table 1. The event rate based on the total calculated SD nucleon cross-section is (422+/-22) events per mt-yr.

*Table 1. RGM-SM Calculated Spin Dependent (SD) WIMP Annihilation Nucleon Cross-Sections*

| Anomaly | Calculated SD Nucleon Cross-Section |
|---|---|
| 0 | $(1.22 \pm 0.40) \times 10^{-42} cm^2 N^{-1}$ |
| 1 | $(2.20 \pm 0.08) \times 10^{-42} cm^2 N^{-1}$ |
| 2 | $(4.46 \pm 0.10) \times 10^{-42} cm^2 N^{-1}$ |
| 3 | $(5.85 \pm 0.13) \times 10^{-42} cm^2 N^{-1}$ |
| Sum | $(1.37 \pm 0.07) \times 10^{-41} cm^2 N^{-1}$ |